\documentclass[aps,pra,twocolumn,groupedaddress,floatfix,showpacs]{revtex4}

\usepackage{color}
\usepackage{units}
\usepackage{graphicx}
\usepackage{bm}% bold math

\begin{document}

%\title{A proposed separation dependent direct-ionization measurement of long range molecules: the extension of electron orbital precession to the molecular case.}

\title{Extending electron orbital precession to the molecular case: Can orbital alignment be used to observe wavepacket dynamics?}

\author{Hugo E. L. Martay}
\email{h.martay1@gmail.com}
\affiliation{Clarendon Laboratory, Department of Physics, University of Oxford, Oxford, OX1 3PU, United Kingdom}
%\affiliation{Clarendon Laboratory, Department of Physics, University of Oxford, Oxford, OX1 3PU, United Kingdom}
\author{David J. McCabe}
%\affiliation{Clarendon Laboratory, Department of Physics, University of Oxford, Oxford, OX1 3PU, United Kingdom}
\author{Duncan G. England}
%\affiliation{Clarendon Laboratory, Department of Physics, University of Oxford, Oxford, OX1 3PU, United Kingdom}
%\author{Melissa E. Friedman}
%\affiliation{Clarendon Laboratory, Department of Physics, University of Oxford, Oxford, OX1 3PU, United Kingdom}
%\author{Jovana Petrovic?}
%\affiliation{Clarendon Laboratory, Department of Physics, University of Oxford, Oxford, OX1 3PU, United Kingdom}
\author{Ian A. Walmsley}
\affiliation{Clarendon Laboratory, Department of Physics, University of Oxford, Oxford, OX1 3PU, United Kingdom}

\date{\today}
\begin{abstract}
The complexity of ultrafast molecular photoionization presents an obstacle to the modelling of pump-probe experiments. Here, a simple optimized model of atomic rubidium is combined with a molecular dynamics model to predict quantitatively the results of a pump-probe experiment in which long range rubidium dimers are first excited, then ionized after a variable delay. The method is illustrated by the outline of two proposed feasible experiments and the calculation of their outcomes. Both of these proposals use Feshbach $^{87}$Rb$_2$ molecules. 
We show that long-range molecular pump-probe experiments should observe spin-orbit precession given a suitable pump-pulse, and that the associated high-frequency beat signal in the ionization probability decays after a few tens of picoseconds. 
If the molecule was to be excited to only a single fine structure state state, then a low-frequency oscillation in the internuclear separation would be detectable through the time-dependent ionization cross section, giving a mechanism that would enable observation of coherent vibrational motion in this molecule.
\end{abstract}


\maketitle

\section{Introduction}

The use of sub-picosecond laser pulses have enabled the observation of rapid processes in the time domain through the use of a pump-probe experiment. A typical experiment involves applying a ``pump'' pulse to a sample, for example a gas of atoms or molecules, to initiate a time dependent process. A second ``probe'' pulse is used after a controllable time delay to make a measurement of the state of the sample at this delay. Because the delay between the pulses may be set very precisely, a time dependent measurement may be made on the sample. Over the last decade, pump-probe experiments have been used for the observation of Rydberg electron wavepackets \cite{wolde_1988, noel_1996, yeazell_1990}, atomic spin-orbit precession \cite{christian_1996}, and molecular vibration \cite{khundkar_1990, baumert_1992, gruebele_1993}.

%The difference in ionization cross section between an excited rubidium atom with different alignments between the excited orbital and the ionizing field direction has been used to observe the orbital precession of the valence electron caused by spin-orbit coupling. 

Recently, the observation of spin-orbit precession in rubidium atoms has been demonstrated \cite{zamith_2000}, and used as a substrate for the coherent control of the pump pulses \cite{chatel_2008}. 

In this paper, the theoretical description of spin-orbit precession is extended to the photoionization of long range Rb$_2$ molecules. It is also shown that the same mechanism that enables the spin-orbit precession to be visible can provide a measurement of the separation of the atoms in the dimer.

Two results are obtained from the extension of the atomic orbital precession to molecular orbital precession. The molecular pump-probe signal contains a high frequency component and a low frequency component, in contrast to the atomic case, when only a high-frequency component is present, at the spin-orbit precession frequency. The high frequency component decays, as a result of the orbital precession rate being dependent on the internuclear separation. The low frequency component is due to a vibrational oscillation of the molecule. 

Both components would be visible in a pump-probe ion signal. The reason is that the Hund's case (c) states change their admixtures of Hund's case (a) states as a function of internuclear separation, and the ionization process effectively measures the Hund's case (a) populations rather than the Hund's case (c) populations due to the contrasting timescales on which the ionization process and spin-orbit coupling operate. This gives an internuclear separation dependent measurement that may be used to observe coherent vibrational motion in the molecule, using a direct ionization scheme.

%The ionization process in long range diatomic molecules is identical to the atomic case since the molecule may be considered to be two separated atoms if they are separated by far enough, and the ionization process is too fast to observe the shifts in the energy levels of the atom pair caused by their mutual interaction. On the other hand, the dynamic process that the pump-probe experiment watches is different in molecules to the atomic case, because the delay between the pump and probe pulse is easily long enough to observe the interaction between the two atoms. 

The calculations of the photoionization dynamics presented here are based on a simple model used to describe the rubidium atom. The model is optimized to recover the atomic energy levels and transition strengths of the rubidium atom. Ionization cross sections and the derived model of above threshold short pulse ionization are presented also. 

Section \ref{model} details the atomic model, and gives the optimization process, section~\ref{cross_sections} give the ionization cross sections of various states as a function of wavelength, section~\ref{molecular_model} gives the molecular model, and sections~\ref{ftl} and \ref{cut_pulse} describe two experiments and their predicted outcomes that show the molecular orbital precession and the molecular vibration in the ion signal.

\section{Atomic model\label{model}}

  The rubidium atom has a core containing 36 electrons and a single valence electron. An effective potential for the valence electron is employed with three parameters. These parameters are fit to several energy levels of the rubidium atom and to the transition dipole moments of several transitions. In calculating transition dipole moments, accuracy is improved by assuming that near the core, the electron is effectively shielded from any incident radiation, and a fourth parameter is used to describe the shielding effect.

 The model Hamiltonian, in atomic units, is given by
\begin{eqnarray}
  V(r) &=& -\frac{Z(r)}{r} - \left\lbrace1-e^{-\left(\frac{r}{\rho}\right)^2}\right\rbrace^2\frac{\alpha_D}{2r^4}.\\
  Z(r) &=& 1+be^{-\left(\frac{r}{a}\right)^c}.\\
  H(t) &=& V(r)+\frac{1}{2}\alpha^2\frac{1}{r}\frac{\partial V}{\partial r}\mathbf{l}\cdot\mathbf{s} + \frac{\nabla^2}{2} + D(\mathbf{r})\varepsilon(t). \label{H_e}\\
 D(\mathbf{r}) &=& (1-e^{-\frac{r}{r_s}})\textbf{r}.
\end{eqnarray}
Here, the first term may be considered to be the Coulomb potential of an effective charge, $Z(r)$ of the core, and the second term is the core polarizability term. The second term is taken from reference \cite{szentpaly_1982}, which gives the parameters for the static electric core dipole polarizability, $\alpha_D$ and core size, $\rho$ the values \unit[8.67]{E$_ha_0^4$} and \unit[2.09]{$a_0$} respectively. The Hamiltonian contains a spin-orbit coupling term. $a$, $b$, $c$ and $r_s$ are the optimised parameters of the model. $r$ is the electron-nuclear separation. $\mathbf{l}$ is the electron orbital angular momentum operator, and $\mathbf{s}$ is the electron spin angular momentum operator. $\alpha$ is the fine-structure constant.

The dipole operator is defined as above in order to include a shielding effect. This can be explained physically by the fact that when the valence electron is in the core region of the atom, the core electrons will shift adiabatically in the presence of an electric field, thus shielding the valence electron from the electric field. It also contains any contribution from core rearrangement that might accompany a transition in the valence electron state that might change the transition strength.

The parameters given here were optimized to give a compromise between the energy levels and the transition strengths. It is possible to optimize the parameters to get more accurate energy levels, but at the expense of the transition dipole moments.

The values for $a$, $b$ and $c$ are chosen to minimize a fitness function, chosen to be
\begin{eqnarray}
 \epsilon &=& \sum_j \left(\frac{E_j(\hbox{model})-E_j(\hbox{experimental})}{h\times\hbox{THz}}\right)^2 \nonumber\\
&&+ \sum_k 1000\log^2\left|\frac{\mu_k(\hbox{model})}{\mu_k(\hbox{experimental})}\right|,
\end{eqnarray}
where the sums are over just the energy levels and transitions selected for optimization. The fitness function contains a weighting between the energy level and the transition dipole accuracies. This value was chosen according to the particular requirements of the experiments presented here: in particular the accuracy of the energy levels had to be accurate to some hundreds of h$\times$GHz.

 Values for the four parameters are given in table~\ref{table_params}, and the experimental observables that went into the optimization are given in table~\ref{table_opt}.

\begin{table}  
\begin{center}
\begin{tabular}{ l c}
	\hline
	\hline
	Parameter & Value\\
	\hline
	$a / a_0$&0.296760\\  %0.2967603123\\
	$b$&      37.68652\\ %37.68652084\\
	$c$&        0.729772\\            %0.7297715504\\
	$r_s / a_0$&1.790893\\                   %1.790893252362286]{$a_0$}\\
	\hline
	
\end{tabular}
\caption{\label{table_params} Optimised parameters for the atomic model presented here.}
\end{center}
\end{table}

\begin{table}  
\begin{tabular}{ l r r r l}
	\hline
	\hline
	Observable & Exp.&Model&\\
	          & \cite{sansonetti_2006}  &&\\
	\hline
  	E(5s) &-1010.025&-1010.027&THz \\
  	E(6s) &-406.468&-406.276&THz \\
  	E(7s) &-221.228&-221.179&THz \\
  	E(8s) &-139.223&-139.206&THz \\
  	E(9s) &-95.687&-95.680&THz \\

	E(5p$_{1/2}$) &-632.918&-632.932&THz \\
	E(6p$_{1/2}$) &-299.065&-299.236&THz \\
  	E(7p$_{1/2}$) &-175.552&-175.663&THz \\
  	E(8p$_{1/2}$) &-115.596&-115.669&THz \\
	E(9p$_{1/2}$) &-81.900& -81.947&THz \\
	E(5p$_{3/2}$) &-625.795&-625.420&THz \\
	E(6p$_{3/2}$) &-296.741&-296.862&THz \\
  	E(7p$_{3/2}$) &-174.500&-174.599&THz \\
  	E(8p$_{3/2}$) &-115.031&-115.101&THz \\
	E(9p$_{3/2}$) &-81.562&-81.608&THz \\
	E(4p$_{5/2}$) &-429.771&-431.418&THz \\
	E(5p$_{5/2}$) &-239.454&-239.750&THz \\
	E(6p$_{5/2}$) &-149.939&-150.004&THz \\
	\hline
	$\mu$ 5s$\longrightarrow$5P$_{1/2}$&2.99&3.016&$q_ea_0$ \\
	$\mu$ 5S$\longrightarrow$6P$_{1/2}$&0.236&0.230&$q_ea_0$ \\
	$\mu$ 5S$\longrightarrow$7P$_{1/2}$&0.0813&0.0762&$q_ea_0$ \\
	$\mu$ 5S$\longrightarrow$8P$_{1/2}$&0.0407&0.0382&$q_ea_0$ \\
	$\mu$ 5S$\longrightarrow$9P$_{1/2}$&0.0252&0.0234&$q_ea_0$ \\
	$\mu$ 5S$\longrightarrow$5P$_{3/2}$&2.98&3.008&$q_ea_0$ \\
	$\mu$ 5S$\longrightarrow$6P$_{3/2}$&0.255&0.265&$q_ea_0$ \\
	$\mu$ 5S$\longrightarrow$7P$_{3/2}$&0.0950&0.0963&$q_ea_0$ \\
	$\mu$ 5S$\longrightarrow$8P$_{3/2}$&0.0504&0.0517&$q_ea_0$ \\
	$\mu$ 5S$\longrightarrow$9P$_{3/2}$&0.0326&0.0333&$q_ea_0$ \\
	$\mu$ 5P$_{3/2}\longrightarrow$6d$_{5/2}$&0.677&0.631&$q_ea_0$ \\

	\hline
	$\Delta E_{SO}$&7.123&7.513&THz\\
	\hline
	\hline
	
\end{tabular}

\caption{\label{table_opt} The experimental observables that went into the optimizations. The energies of 18 levels were optimized, as were eleven transition dipole moments.}
\end{table}
%me}P,M_l=0\rangle|^2}{2E_{5S}-E_{nP}-E_{n^{\prime}P}},\]

The model is much simpler than multi-electron models which require configuration interaction calculations or self consistent methods to be applied, and allow calculations to be performed much more easily. When compared to other single-electron models, such as reference~\cite{marinescu_1994}, the model presented here recovers remarkably accurate transition strengths and ionization cross sections, as well as treating spin-orbit coupling accurately. These are essential for estimating the visibility of the spin-orbit precession in an ion signal, and so the model is well suited to the applications presented here. 

The numerical approach to solving the Schr\"{o}dinger equation for Eqn.~\ref{H_e} is given in the Appendix.

\begin{figure}
 \centering
 \includegraphics[width=\columnwidth, angle=0]{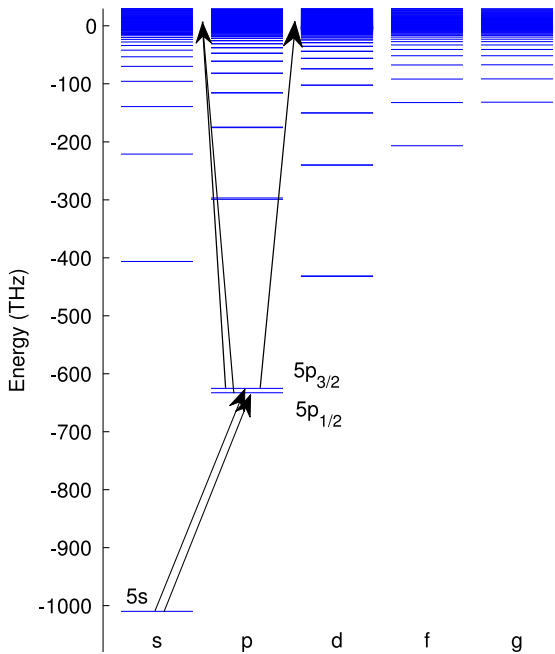}
 \caption{\label{level_diagram}(Color online) The energy levels of the rubidium atom. The atoms in the experiments discussed here are initially in the ground state, 5s. A short pulse transfers them to 5p states. An ionizing pulse then transfers population to continuum states with either s or d symmetry. If both accessible 5p states are populated and the ionizing pulse is short enough, then interference effects cause the precession of the orbital angular momentum about the total angular momentum to be visible in the ion signal.}
\end{figure}

\section{Atomic photoionization cross section \label{cross_sections}}
	
Using this atomic model, the photon energy dependent photoionization cross section of the ground state is calculated for verification as described in the Appendix.

\begin{figure}
 \centering
 \includegraphics[width=\columnwidth, angle=0]{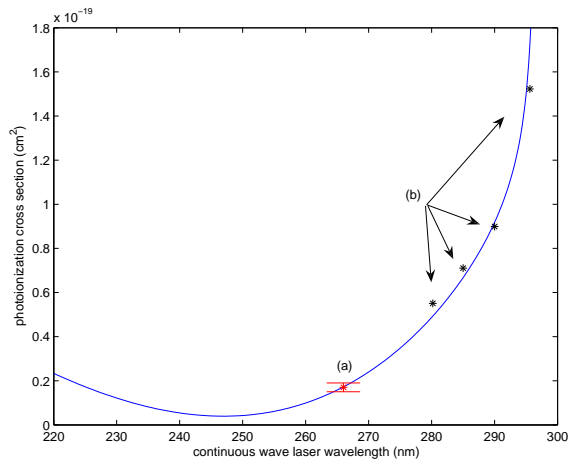}
 \caption{\label{5sPI}The continuous-wave photoionization cross section as a function of wavelength for the ground state rubidium atom. The blue line is the model's predictions. The point (a) shows the result from reference~\cite{lowell_2002}, including error. The two results are in agreement. The points (b) shows the results from reference.~\cite{suemitsu_1983}. These are a relative magnitude measurement and so have been scaled appropriately.}
\end{figure}

	The continuous wave photoionization cross section as a function of photon energy for the ground state is shown in Fig.~\ref{5sPI}. The model agrees with experiment \cite{lowell_2002, suemitsu_1983} for the photoionization of the ground state, but overestimates cross sections for the photoionization cross section of the 5P state by around 20\% when compared to experiment \cite{dinneen_1992}. This could be for a variety of reasons. Firstly, the error bars on the experimental result are wide, and it is possible that the true value is quite close to the calculation presented here. Secondly, the photionization cross sections of the 5p states depend on the transition dipole moments from p orbitals to d orbitals. Due to the scarcity of experimental transition dipole moments from p to d orbitals, the model was only optimized to recover a single p to d transition. Therefore it is quite likely that although the s and h{p} electron wavefunctions are quite well recovered, the d orbital wavefunctions could be less accurate, and this affects the p orbital ionization cross sections. 

	For short-pulse photoionization, the ionization probability may be expressed as the magnitude of the final wavefunction when projected onto the scattering electronic states:
\begin{eqnarray}
 	P(\hbox{ion}) \simeq  |\wp_{\hbox{ion}} U \psi_{\hbox{i}} |^2,
\end{eqnarray}
	where $U$ is the propagator from before the start of the pulse to after the end of the pulse, $\wp_{\hbox{ion}}$ projects onto only ionised electronic states, and $\psi_{\hbox{i}}$ is the initial state. Equivalently, this can be expressed as the expectation value of an operator, $\hat{M}$.

The atom has six 5P excited states: two spin states multiplied by the three orbital angular momentum projection quantum numbers: -1, 0 and 1. The axis of projection will be chosen to be the electric field axis of the ionizing electric field. In the fine structure basis, the 5P$_{1/2}$ state has 2-fold degeneracy and the 5P$_{3/2}$ state has 4-fold degeneracy. 
	Any operator acting on this basis may be expressed as a six by six matrix, but the reversal symmetry ( $m_{\ell}\rightarrow-m_{\ell}$, $m_s\rightarrow-m_s$) means that the states with positive total angular momentum are never coupled to states with negative angular momentum, and so we can limit the discussion to the positive to positive matrix elements in the understanding that the negative to negative elements are the same. More generally, neither the electric field nor the spin-orbit coupling changes the total projected angular momentum of the atom, so the four possible values, -3/2, -1/2, +1/2 and +3/2 remain separate: Nothing in this model couple between values for this quantum number. The ionization operator will therefore be block diagonal with each block representing a different total angular momentum.

The three positive total angular momentum basis states are
\begin{eqnarray}
 |j=1/2, m_j=1/2\rangle,\\
 |j=3/2, m_j=1/2\rangle,\\
 |j=3/2, m_j=3/2\rangle.
\end{eqnarray}
	If an arbitrary 5P wavefunction $|\psi\rangle$ is represented by a vector $\mathbf{a}$:
\begin{eqnarray}
 	|\psi\rangle &=& a_1 |j=1/2, m_j=1/2\rangle \nonumber \\ \nonumber
			&&+ a_2|j=3/2, m_j=1/2\rangle\\ \nonumber
			&&+ a_3 |j=3/2, m_j=3/2\rangle, \nonumber
\end{eqnarray}
	then the ionization operator, $\hat{M}$, whose expectation value gives the ionization probability is a three by three Hermitian matrix.

	As an example, the ionization operator whose expectation value gives the ionization probability for a Gaussian probe pulse, centred at \unit[650]{THz}, with a full width at half maximum of \unit[18]{THz}, and with a total fluence of \unit[0.3]{Jm$^{-2}$} is given as
\begin{eqnarray}
\hat{M}= \left(\begin{array}{c c c}
  0.997    &   -0.223       &                 0\\
        -0.223     &   1.218     &                   0\\
                         0        &                0    &    0.817\\
       \end{array}\right)\times10^{-3}.
\end{eqnarray}

A basis change can be made to the $s,m_s, \ell,m_{\ell}$ basis, so that the 5P state is expressed as 
\begin{eqnarray}
 	|\psi\rangle = b_1 |m_\ell=0, m_s=+1/2\rangle \nonumber \\ \nonumber
	+ b_2|m_\ell=1, m_s=-1/2\rangle \\ \nonumber
	+ b_3 |m_\ell=1, m_s=+1/2\rangle. 
\end{eqnarray}

In this basis, the ionization matrix takes the form 
\begin{eqnarray}
\hat{M}= \left(\begin{array}{c c c}
 1.355    &   0.0297   &                    0\\
        0.0297    &    0.860     &                 0\\
                       0                  &      0    &    0.817\\
       \end{array}\right)\times10^{-3}.
\end{eqnarray}

The small off-diagonal elements show that the ionization process effectively measures the populations in the $m_\ell$ basis. The physical reason for this is that the pulse is shorter than the spin-orbit interaction time, and so the electron spins may be neglected, leaving the differing $\ell,m_\ell$ quantum numbers as the elements that affect the ionization probability. The difference in ionization probability between the two $m_\ell=1$ states is due to the subtly different spatial wavefunction of the $j=3/2$ and $j=1/2$ states.%see ref.~[cite fine structure anomaly review]). 

The ionization matrix in this example, and in all the cases presented, was calculated by propagating the three initial states under the influence of the electric field and using their ionization probability and the inner product between the resulting ionized wavefunctions to infer matrix elements. Although the field used here is in the weak field regime, this method of parameterizing the effect of the field applies to any field. 

The ionization process will be approximated in the molecular simulations as 
\begin{eqnarray}
 \hat{M}=&&|m_\ell = 0\rangle 1.36\cdot10^{-3}\langle m_\ell = 0 |\\
	&+& |m_\ell = 1\rangle 0.84\cdot10^{-3}\langle m_\ell = 1 |.\label{M_ion}\\
\end{eqnarray}
The approximation is correct to within a few percent, which is also roughly the error in transition dipole moments of the atomic model.

\subsection{Ionization operator dependence}

\begin{figure}
\begin{center}
 \includegraphics[width=\columnwidth, angle=0]{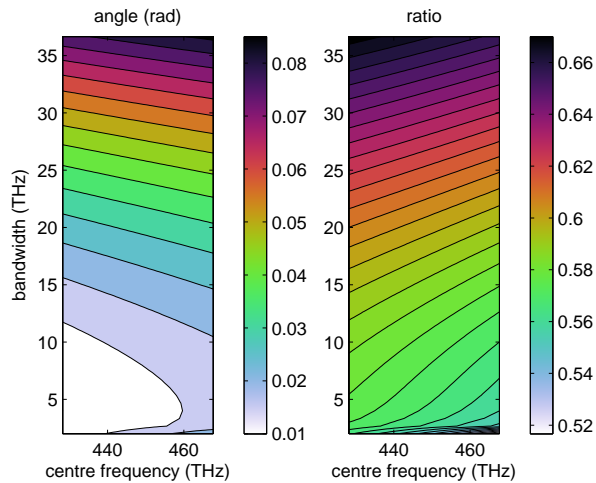}
 \caption{\label{ionization_parameterised}The ionization operator is a function of the probe pulse parameters. The operator has a principle axis defined as the state that is most easily ionized. The angle between this and the $|m_\ell=0\rangle$ state is shown in the first panel. The ratio of ionization probabilities for the $|m_\ell=1\rangle$ compared to the $|m_\ell=0\rangle$ state is shown in the second panel. The result is that both quantities have a weak dependence on the probe pulse, but narrower bandwidth probes are more sensitive to orbital alignment.}
\end{center}
\end{figure}

	The framework described above is used to find the ionization parameters as a function of pulse bandwidth and centre frequency. The two quantities that determine whether the ionization operator is selective for one orbital orientation are the angle between the operator's principle axis and the $|m_\ell=0\rangle$ axis, and the ratio of ionization probabilities for the $|m_\ell=1\rangle$ and $|m_\ell=0\rangle$ states. These two values are shown in Fig.~\ref{ionization_parameterised} as a function of probe wavelength and bandwidth. The ionization operator can be seen to vary slowly with laser parameters over these intervals, becoming more orbital-alignment selective at smaller probe bandwidths.  

\begin{figure}
 \centering
 \includegraphics[width=\columnwidth, angle=0]{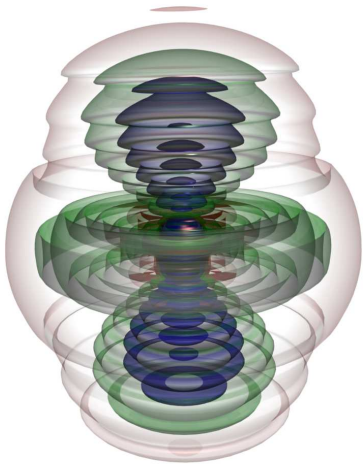}
 \caption{\label{atomic_state_illustratio}The wavefunction of the valence electron at the peak of a short ionizing laser pulse. Contours of probability density multiplied by the electron-core separation squared are chosen at intervals to highlight the exiting electron's wavefunction, which is a superposition of s and d waves. Because of the short timescale, the electron has not had time to travel far from the core. The shelled structure is due to the electron being emitted at a specific phase of the incident radiation. The electric field axis is aligned vertically.}
\end{figure}

\section{Atomic spin-orbit dynamics}
The spin-orbit dynamics of the first excited state of the rubidium atom are of most interest here due to the ease of production of the state. The two levels in question are illustrated in Fig.~\ref{level_diagram}.

The atom has a single valence electron, giving it an electron spin of $\frac{1}{2}$. In the first excited state, the 5P state, the electron orbital has one unit of angular momentum. This gives the atom 6 allowed angular momentum states. Two pairs of these are coupled by the spin-orbit coupling: $|m_\ell = 0, m_s = \pm \frac{1}{2}\rangle$ is coupled to $|m_\ell = \pm1, m_s = \mp \frac{1}{2}\rangle$. Since the atom is transferred by a short laser pulse to the $|m_\ell = 0\rangle$ state, where the quantization axis is chosen along the electric field of the exciting laser pulse, the spin-orbit coupling causes an oscillation between the $|m_\ell = 0\rangle$ and $|m_\ell = 1\rangle$ states. Since the $|m_\ell = 0$ state is about 60\% easier to ionize than the $|m_\ell = 1\rangle$ state, this gives the pump-probe signal of the atom an oscillation at the spin-orbit frequency of \unit[7.123]{THz}. 

\begin{figure}
 \centering
 \includegraphics[width=\columnwidth, angle=0]{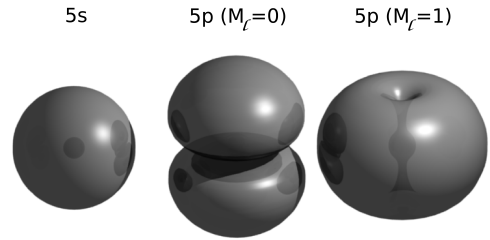}
 \caption{\label{atomic_state_illustration}The rubidium ground state and first excited states. In the experiments discussed, the atom starts off in the ground state (5s) and is transferred to the first excited state. If the transition is instantaneous, then the excited population will be in the $m_{\ell} = 0$ state (centre). This is not an eigenstate of the full Hamiltonian since it is spin orbit coupled to the $m_{\ell}=1$ state (right). The result is that the electron orbital oscillates between the $m_{\ell} = 0$ and $m_{\ell} = 1$ states. Since the two states have different ionization probabilities when an ionizing pulse is applied, the probability of ionizing the atom oscillates. The surfaces are a contour of $\rho(\mathbf{r})|r|^2$.}
\end{figure}

The initial state and the $|m_\ell = 0\rangle$ and $|m_\ell = 1\rangle$ states are shown in Fig.~\ref{atomic_state_illustration}.

\section{Extension to molecular case\label{molecular_model}}
	The calculations can be extended to the molecular case by increasing the 8 atomic states (2 ground, 6 excited) to 15 molecular states (3 ground, 12 excited), which make up all the accessible electronic states of the molecule, assuming it only absorbs a single photon and is initially in a triplet state. The internuclear separation is added as a continuous variable,  and the angle the internuclear axis makes with the electric field of the laser pulses is added as a constant parameter since it does not change appreciably over the timescale of the experiment. The Schr\"{o}dinger equation is thus converted from a discrete 8 by 8 matrix equation into a one dimensional partial differential equation with 15 channels. The details of the model are given in a previous publication \cite{martay_2009}.

	The ionization step is modelled by projecting the molecular state onto the separated atomic states and taking the expectation value of the ionization matrix (Eqn.~\ref{M_ion}) as the ionization probability.

\section{Experiments}
	The application of the atomic model to the study of the direct photoionization of long range molecules is demonstrated by considering two experiments. 

	The two experiments outlined here are both possible, and an effort has been made to ensure they are feasible with current technology. Both have the same initial state which can be prepared using a ramped magnetic field over a Feshbach resonance. In both experiments, a pump pulse is applied, then after a delay, a probe pulse with the same polarization is applied. The probe pulses are identical, but the pump pulses are different. The polarization was chosen parallel to the magnetic field used to associate the molecules.

In the first experiment, the pump pulse is a Fourier limited Gaussian pulse whose bandwidth spans both the transition from the ground state, 5S, to the 5P$_{1/2}$ and to the 5P$_{3/2}$ state. This maximises the visibility of the spin-orbit precession. In the second experiment, the same Gaussian pump pulse is spectrally cut to remove intensity at the 5S --- 5P$_{3/2}$ transition. This prevents the preparation of a precessing orbital. As a result, the only mechanism by which the orbital can change orientation is through the interaction of the two atoms in the molecule. Since this depends on internuclear separation, the oscillation in the internuclear separation coordinate becomes apparent in the pump-probe signal.

\subsection{Initial state\label{initial}}
	The initial state for each experiment is prepared using a magnetic field sweep across a Feshbach resonance in a cold $^{87}$Rb gas with a high phase space density in order to associate atom pairs to loosely bound molecules.

	The ground electronic state rubidium dimer has a discrete number of bound states plus several continua of scattering states, each with a different electron and nuclear spin state. The energies of the bound and continuum states change as a function of magnetic field and spin state. The dynamics and molecular properties of long range dimers bound by less than a few h$\times$GHz is dictated by the hyperfine and Zeeman interactions. Several reviews \cite{kohler_2006} and other publications are concerned with the behaviour of such molecules, which constitute a major area of study in their own right. 

The use of a Feshbach resonance to create molecules was demonstrated in 2002 \cite{donley_2002}. Work had already been done on using a magnetic field to control atom-atom interactions both theoretically and experimentally \cite{tiesinga_1993, inouye_1998}. This mechanism is the mechanism for creating the initial state in the work here. An essay by D. Kleppner presents a fuller picture of the use of Feshbach resonances in this context \cite{kleppner_2004}. Magnetic sweep association of the species under the conditions studied here has been demonstrated experimentally \cite{thalhammer_2006}.

\begin{figure}
 \centering
 \includegraphics[width=\columnwidth, angle=0]{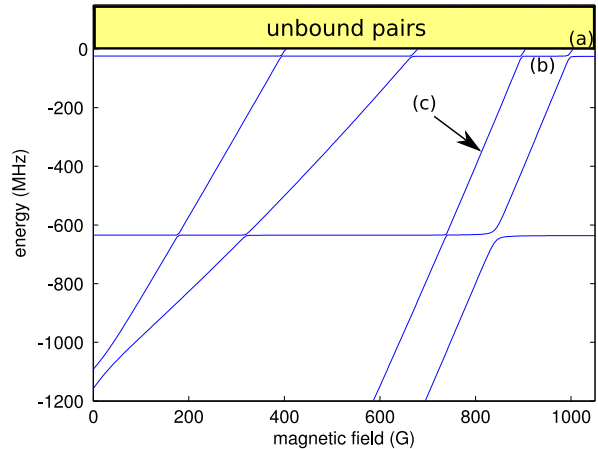}
 \caption{\label{feshbach_diagram}The highest bound state energies of the rubidium dimer as a function of magnetic field. For the experiments outlined here, the atom pairs are initially unbound at a magnetic field of greater than \unit[1007]{G} (a). The magnetic field is swept down to \unit[940]{G} (b) for the first experiment, and to \unit[850]{G} (c) for the second experiment. This prepares a suitable initial state for each experiment.}
\end{figure}

	Two interacting atoms are prepared in their high-field seeking $f = 1$ state. The spin-spin interaction between the two atoms is neglected, which constrains the atom pair to stay in the $m_f = 2$ state, since the remaining hyperfine, Zeeman, and electronic interactions do not break the rotational symmetry of the electron plus nuclear spins about the magnetic field axis. For numerical reasons, the internuclear separation is constrained to be in a box. The highest 9 vibrational states and lowest 191 box states for each of the 5 accessible spin states are taken as a basis set. The Hamiltonian, consisting of the vibrational energies, the electron and nuclear Zeeman energies, and the hyperfine interaction, was diagonalised for each magnetic field.  The energy levels of the dimer as a function of magnetic field for this model are shown in Fig.~\ref{feshbach_diagram}. 

	By following a single state adiabatically from above \unit[1007]{G}, where it is a box state, to \unit[940]{G}, where it is bound by 24 h$\times$MHz, or to \unit[850]{G} for the second experiment, a suitable initial state is prepared. 

	For the \unit[940]{G} state, the state is well represented by the highest single channel vibrational state (whether singlet or triplet) multiplied by a spin state $|\psi\rangle$ which is 74\% $|S=1, M_S = -1, I=3, M_I=3\rangle$, \newline 12\% $|S=1, M_S = 0, I=3, M_I=2\rangle$, and 12\% $|S=0, M_S = 0, I=2, M_I=2\rangle$. The superposition is coherent, but the phases between these three states are not measured by the experiment, since the rest of the experiments take place on a timescale too short to change the nuclear spins and so the nuclear spins effectively measure the initial state of the electron spin. For this state, the singlet population is neglected. There are two justifications for this. The first is that it is only 12\% of the population, and the second is that it is expected to behave similarly to the triplet population: It will spin-orbit precess at first, but molecules at different internuclear separations will spin-orbit precess at different rates, so the visibility of the precession will wash out.

	For the \unit[850]{G} state, the state is well represented by the fifth highest triplet state with 70\% in the \newline $|S=1, M_S = 1, I=3, M_I=1\rangle$ state, and 28\% in the $|S=1, M_S = -1, I=3, M_I=3\rangle$ state.

% The spatial part of the triplet wavefunction is shown in Fig.~\ref{initial state wavefunction}.

\begin{figure}
 \centering
 \includegraphics[width=\columnwidth, angle=0]{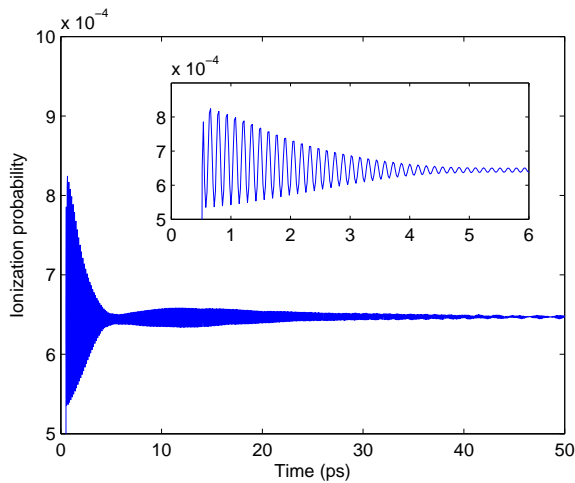}
 \caption{\label{pump_probe_signa_uncut}The ionization probability of a long-range molecule prepared as described in section~\ref{initial} at \unit[940]{G} in a pump-probe experiment where the molecule is excited to the states associated with the 5S + 5P asymptote and then ionized directly. The pump and probe are sub-picosecond pulses. This is the calculated experimental outcome for experiment 1, where the molecules are put in a superposition of fine structure states in order to cause spin-orbit precession. The spin-orbit precession is clearly visible in the signal here.}
\end{figure}
\begin{figure}
 \centering
 \includegraphics[width=\columnwidth, angle=0]{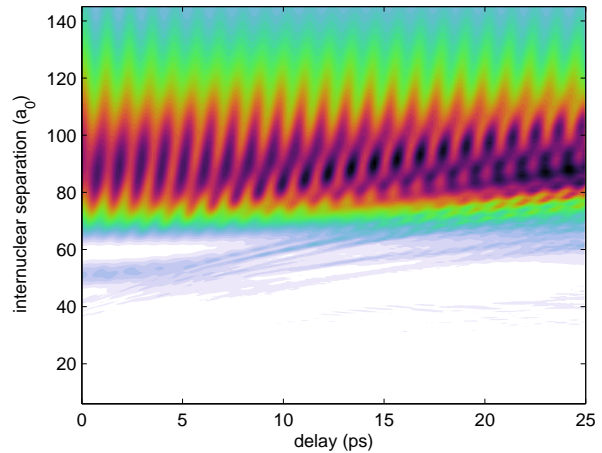}
 \caption{\label{pump_probe_signal_uncut}The relative number of molecular ions yielded as a function of internuclear separation for a long-range molecule prepared as described in section~\ref{initial} in a pump-probe experiment where the molecule is excited to the states associated with the 5S + 5P asymptote and then ionized directly. The angle between the internuclear axis and the electric field is 1.4 radians. The contribution is shown as a function of internuclear separation and time. This is the population density multiplied by ionization probability for experiment 1. It shows the spin orbit precession, and how the contributions from different internuclear separations become dephased leading to the signal washing out. The figure has been deliberately undersampled to increase the apparent oscillation period to around \unit[1]{ps} rather than \unit[1/7]{ps}.}
\end{figure}

\subsection{Experiment 1: Broadband uncut pump\label{ftl}}
	In this experiment, we wish to observe spin-orbit precession in the long range Rb$_2$ dimer. The molecule is prepared as described in section~\ref{initial}. This initial state is very loosely bound, which increases the dephasing time for the spin-orbit precession, allowing a clearer signal. The molecules are excited with a Gaussian pump probe with a centre wavelength of \unit[375]{THz} and a full width at half maximum intensity of \unit[30]{THz}. After a variable time delay, the molecules are ionized by a second Gaussian laser pulse, centred at \unit[650]{THz}, with a full width at half maximum of \unit[18]{THz}, and with a total fluence of \unit[0.3]{Jm$^{-2}$}. The ionization probability as a function of delay is measured by counting the number of molecular ions produced. 

	The experiment is modelled by taking the initial state, propagating it under the influence of the first pulse. At each timestep, the expectation of the ionization operator can be taken, which gives the ionization probability of the molecule at the specified delay. The calculation is repeated for 21 different angles between the internuclear axis and the electric field and averaged. This ionization probability is given in Fig.~\ref{pump_probe_signa_uncut}. The contribution to this signal as a function of internuclear separations for one angle is given in Fig.~\ref{pump_probe_signal_uncut}.

	The spin-orbit precession is initially visible as a \unit[7]{THz} oscillation, with a good signal to background, but then decays due to the interaction of the two atoms.

\begin{figure}
 \centering
 \includegraphics[width=\columnwidth, angle=0]{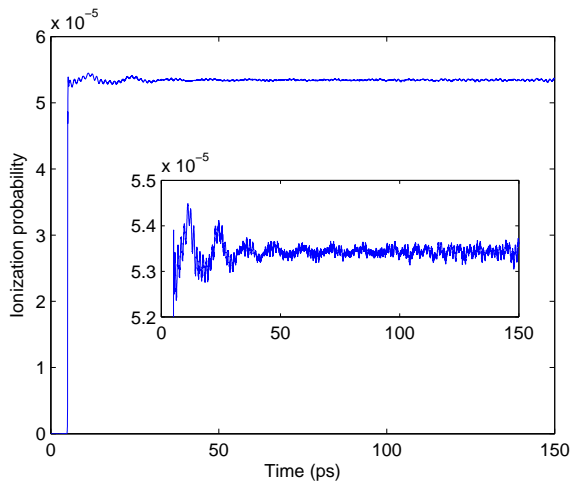}
 \caption{\label{pump_probe_signa_cut}The ionization probability of a long-range molecule prepared as described in section~\ref{initial} at \unit[850]{G} in a pump-probe experiment where the molecule is excited to the states associated with the 5S + 5P asymptote and then ionized directly. This is the signal for the second experiment. The pump pulse is spectrally cut to suppress the spin-orbit precession. As a result, and as a result of the different initial state, the molecular dynamics are visible as a long period contribution to the signal.}
\end{figure}

\begin{figure}
 \centering
 \includegraphics[width=\columnwidth, angle=0]{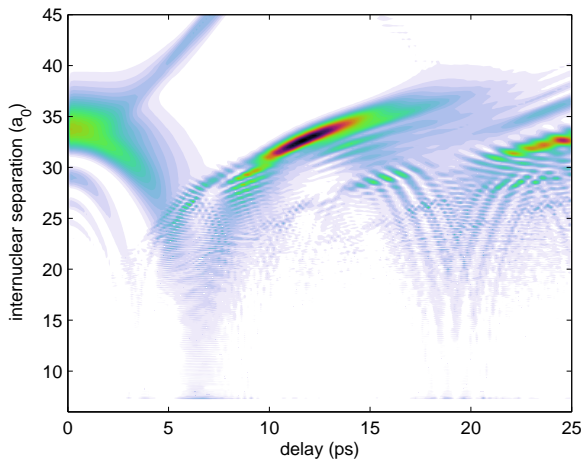}
 \caption{\label{pump_probe_signal_cut}The contribution to the total ion yield of a long-range molecule prepared as described in section~\ref{initial}, as for Fig.~\ref{pump_probe_signa_uncut}, but for experiment 2. It shows the internuclear separation oscillations clearly. The admixture of excited states that ionize easily change with internuclear separation, and this results in a time-dependent ion signal.}
\end{figure}

\subsection{Experiment 2: Spectrally cut pump\label{cut_pulse}}
 	In this experiment, the aim is to observe the internuclear oscillations rather than spin-orbit precession. As a result, a different initial state is chosen that has the fifth highest triplet state occupied rather than the highest. This shifts the molecular population to separations where the ionization process is more sensitive to separation, and where the population oscillates more uniformly. The pump pulse is spectrally cut to suppress the orbital precession. The probe is the same as for experiment 1. The experiment is modelled as before, and the expected ionization probability is given in Fig.~\ref{pump_probe_signa_cut}. The contributions to this signal are shown in Fig.~\ref{pump_probe_signal_cut}.

	The spin-orbit precession is suppressed as hoped. A persistant long period oscillation may be observed mixed with a short period oscillation. The long period signal is due to the oscillation in the internuclear separation. The short period signal is due to spin-orbit precession. 
\subsection{Discussion}
	The first experiment would demonstrate that it is possible to observe spin-orbit precession in long-range molecules. One application for this is to verify which molecular vibrational state has been prepared by a magnetic field sweep, since the speed with which the oscillation in the ion signal washes out depends on the vibrational state or states that the molecules are in.
	
	The second experiment is similar to the first experiment discussed in reference~\cite{martay_2009}. In the earlier work, it was shown that a coherently oscillating excited state wavefunction could be formed in long-range molecules. It was stated that if a position sensitive measurement could be made, then the molecular dynamics could be observed. In this paper, the result is extended by showing that a position dependent measurement can indeed be made, using orbital alignment. The initial state has to be modified from the earlier work, since it is difficult experimentally to use an above threshold ionization laser on a gas in a magneto-optical trap due to atomic ionization. In this way, the pump-probe experiment can now be modelled from initial state preparation to the creation of the molecular ions. The observation of wavepacket dynamics in the excited state of long range molecules would be a step towards the control and stabilization of such molecules, and would be a valuable contribution to the field of ultracold chemistry.
 
\section{Conclusions}

	A simple atomic model is presented that is accurate enough to model spin-orbit precession and photoionization. A method for extending atomic spin-orbit precession calculations to the case of long range molecules is presented. This is demonstrated by the calculation of the ionization signal for two pump-probe experiments. A consequence of this is that it is shown that long range molecules can exhibit spin-orbit precession, but that it washes out. A second consequence is that it is shown that the mechanism that causes spin-orbit precession to affect the ionization signal also causes an internuclear separation oscillation to affect the ionization signal, allowing pump-probe experiments to observe coherent wavepacket behaviour in ultracold long range molecules.

\begin{acknowledgments}
	The authors wish to acknowledge discussion and advice from Thorsten K\"{o}hler, Jordi Mur-Petit and Jovana Petrovi\u{c}.
Financial support is acknowledged from the EPSRC, grant number EP/D002842/1.
\end{acknowledgments}

\section*{APPENDIX}
%\section{Numerical determination of the atomic energy spectrum}
	The eigenstates and eigenenergies of the Schr\"{o}dinger equation are found in the basis defined by the quantum numbers of total angular momentum $j=l+s$, the projected total angular momentum $m_j = m_l + m_s$, and the electron orbital angular momentum $l$ --- a basis for which equation \ref{H_e} is diagonal. The lowest 100 energy levels for the s,p,d,f and g orbitals are used. The calculations are performed in a spherical box of \unit[1600]{$a_0$}. This ensures that the lowest 100 energy levels contain scattering states with an energy range that covers the possible free electron energies in the experiments discussed here. The finite spacing of the scattering states puts an upper bound on the duration of any ionizing pulse that can be accurately modelled. However, all the pulses used here are short enough that the scattering states included in the basis set provide an accurate representation of the continuum. The energy levels were obtained using fourth order Runge-Kutta.

	The time-dependent Schr\"{o}dinger equation was solved using a matrix representation of the propagator for each timestep, chosen to be shorter than the time variation in the Hamiltonian. For weak fields, a 1st or 2nd order perturbative expansion may be used.   

	Continuous-wave photoionization cross sections can be calculated in this framework using the equation
\begin{eqnarray}
 \sigma \simeq \frac{4\pi}{\sqrt{2}} LE^{-\frac{3}{2}}\omega \sum_j\mu_j^2,
\end{eqnarray}
	where $\mu_j$ is the transition dipole moment from the initial state to the final scattering state at energy $E$ normalized to a box of length $L$, and $\omega$ is the angular frequency of the incident linearly polarized light. The sum is over the various electronic configurations of the scattering state: the different allowed orbital and spin angular momenta of the exiting electron. The equation becomes exact in the limit that $L\rightarrow\infty$.

\end{document}